\documentclass[a4paper]{article}

\usepackage{amssymb,amsfonts}       

\usepackage{epsfig}
 
\usepackage{amsmath}
\usepackage{rotating}
 
\usepackage[T2A]{fontenc}
\usepackage[koi8-r]{inputenc}
\usepackage[english,russian]{babel}
 
\begin{document}

\vspace{1cm}

\centerline{\Large \bf An oscillating universe model
based on chromoelectric fields}

\vspace{0.2cm}
\centerline{V.\,N. Yershov}
\vspace{0.3cm}
\centerline{\small Mullard Space Science Laboratory,}
\centerline{\small Holmbury St.Mary, Dorking RH5 6NT, UK}
\centerline{\small vny@mssl.ucl.ac.uk}
\vspace{0.7cm}

\centerline{Abstract}
\vspace{0.2cm}
\hrule
\vspace{0.2cm}
\noindent
An oscillating universe model is discussed, in which the 
singularity of the initial state of the universe 
is avoided by postulating an upper limit on spacetime 
curvature. This also results in the devising of the 
simplest possible structure -- a primitive particle 
(usually called preon), which can be considered as 
the basic constituent of matter that have no properties 
except for the property of carrying a unit chromoelectric
charge. The SU(3)$\times$U(1)-symmetry of its field 
results in the emergence of a unique set of structures 
reproducing exactly the observed variety of the 
fundamental fermions and gauge bosons. The discussed
scheme allows finding answers to many fundamental questions
of the standard particle physics and cosmology 
based on a very few primary constituents.
\vspace{0.2cm}

\hrule
\vspace{0.2cm}
\centerline{(in Russian)}

\title{\bf Модель осциллирующей вселенной на основе хромоэлектрических
полей}
\author{Ершов В. Н.\\
{\small Маллардовская лаборатория космических исследований} \\
{ \small Holmbury St.Mary, Dorking RH5 6NT, UK} \\
{ \small vny@mssl.ucl.ac.uk}}

\date{}

\maketitle

\begin{abstract}
{\small 
Обсуждается модель осциллирующей вселенной, в которой 
отсутствует сингулярность начального состояния
за счет постулирования ограничения на кривизну 
пространства-времени. Это также приводит к 
пространственно-временной структуре, которая может 
рассматриваться в качестве простейшего
компонента модели композитных элементарных частиц 
(обычно называемого преоном) и у которой отсутствуют 
какие-либо свойства, за исключением свойства обладания 
единичным хромоэлектрическим зарядом. Симметрия 
SU(3)$\times$U(1) поля этой простейшей частицы приводит к 
формированию набора структур, единственным образом 
воспроизводящего наблюдаемый в природе набор фундаментальных
фермионов и калибровочных бозонов. Обсуждаемая схема 
позволяет ответить на многие проблемные вопросы 
стандартной модели элементарных частиц и космологии 
опираясь всего лишь на несколько исходных принципов.
} 
\end{abstract}

\newcommand{\rum}{\rule{0.5pt}{0pt}}
\newcommand{\rub}{\rule{1pt}{0pt}}
\newcommand{\rim}{\rule{0.3pt}{0pt}}
\newcommand{\numtimes}{\mbox{\raisebox{1.5pt}{${\scriptscriptstyle \times}$}}}

\newcommand{\abs}[1]{\mid\negmedspace#1\negmedspace\mid}

\section{Введение}
\label{introduction}

В последние годы старая идея осциллирующей вселенной 
\cite{lemaitre33} получила новое развитие 
после публикации серии работ 
Штейнхардта и Турока \cite{steinhardt02a},
в которых описывается модель, использующая 10-мерные сталкивающиеся браны.
В этой модели преодолены основные трудности предыдущих моделей
осциллирующей вселенной, такие как проблема неубывающей энтропии 
\cite{markov84} или отсутствие реалистичного физического механизма, 
приводящего к расширению вселенной после каждого очередного 
ее сжатия \cite{guth83}.
Однако в этой модели требуется, чтобы вселенная имела десять 
пространственных измерений, что пока не подтверждается наблюдениями. 
Поэтому здесь мы вернемся к осциллирующей модели с тремя пространственными 
измерениями, в которой пространство и материя отождествляются
между собой, как это в свое время было предложено 
Уилером \cite{wheeler62}.
А именно, мы будем рассматривать частицы материи в виде стабильных 
конфигураций движущегося многообразия (пространства), 
полагая, 
что геометрия является существенно динамическим понятием 
и что именно геометрия пространства-времени и определяет 
свойства материи \cite{rovelli04}.

Мы рассмотрим здесь топологический аспект 
проблемы и попытаемся за счет этого выяснить природу космологической 
сингулярности. А именно, предположив, что структура начального состояния 
вселенной имела топологию трехмерной бутылки Клейна,
мы избавимся от этой сингулярности.
Этот же подход поможет нам справиться также и с
проблемой сингулярности микро- и 
макроскопических объектов, таких как элементарные частицы 
и сколлапсировавшие звезды. 

При этом мы предположим, 
что общая теория относительности справедлива вплоть 
до расстояний, соответствующих планковской длине, 
следуя идеям Маркова \cite{markov82} 
о том, что в природе должно существовать ограничение на
кривизну пространства.
При реализации данной идеи мы заменим сингулярность 
на топологическую особенность многообразия,
используемого для представления пространства-времени. 
Такая замена возможна, поскольку известны 
решения уравнений поля Эйнштейна, свободные от 
сингулярностей, например, мост Эйнштейна-Розена \cite{einstein35}. 
Кроме того, существует много других моделей, 
в которых топологические особенности 
рассматриваются в качестве частиц материи или систем 
взаимодействующих частиц \cite{finkelstein59}. 
Здесь ма рассмотрим возможность того, 
что такие системы действительно имеют 
отношение к внутренним структурам кварков и лептонов (фундаментальных
фермионов). Другими словами, в рамках данного подхода 
мы будем рассматривать фундаментальные
фермионы в качестве композитных частиц. 

В большинстве существующих композитных моделей кварки и
лептоны представляются в виде комбинаций нескольких (обычно 
двух или трех) более простых
частиц, называемых преонами \cite{dsouza92}. 
В настоящее время преонные модели перестали пользоваться
популярностью из-за того, что
они сталкиваются с многочисленными теоретическими трудностями, 
такими как, например, проблема масс преонов. 
Так, из экспериментов по рассеянию частиц известно, что
гипотетическая шкала расстояний композитности элементарных частиц
соответствует расстояниям, меньшим чем $10^{-18}$~м.
Неопределенность импульса частицы, заключенной в таком небольшом объеме 
пространства, составляет примерно 200~ГэВ, 
что значительно превышает массы кварков первого семейства.
Данную трудность можно преодолеть путем постулирования новой
силы, которая на много порядков превосходила бы силу сильного
взаимодействия.  
За счет такой \glqqгиперсилы\grqq \, преоны могут находиться в связанном
состоянии внутри кварков, причем огромная энергия их импульсов 
компенсируется их большой энергией связи (дефектом массы).
На наш взгляд, данный подход является весьма многообещающим, и
здесь мы будем придерживаться именно этого подхода.
 
Но главным отличием представляемой здесь модели от других
преонных моделей будет то, что здесь мы сократим до предела
число возможных типов базовых частиц.
Как мы уже упомянули, все предыдущие композитные модели
объясняют наблюдаемое в природе разнообразие элементарных частиц
за счет различных комбинаций определенного набора преонных типов,
существенно меньшего, по сравнению с числом типов 
фундаментальных частиц стандартной модели  \cite{pati74}. 
Однако очевидно, что даже это уменьшенное число
фундаментальных типов не может решить проблему, поскольку,
заменяя один набор базовых элементов на другой мы все-равно 
должны искать объяснение происхождения этого нового 
набора базовых элементов и ответить на вопрос о том, 
{\sl почему} частицы в этом наборе различаются между собой.
И только модель основанная на  
{\sl одном единственном} базовом элементе может иметь
смысл. Как-раз это мы здесь и собираемся предложить.
А именно, мы предположим, что существует единственный тип 
простейших частиц (преонов),
не обладающих никакими характеристиками или квантовыми числами
кроме электрического и цветового зарядов. То есть, преоны в 
предлагаемой модели будут преставлены единичными зарядами 
с симметрией SU(3)$\times$U(1). Причину возникновения 
разницы между преонами с противоположными зарядами 
мы попытаемся объяснить в следующем параграфе.

\section{Базовая частица}
\label{primitiveParticle}

\begin{figure}[htb]
\begin{turn}{-90}\epsfig{figure=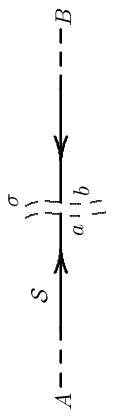,width=3.0cm}\end{turn}
\put(100,-0.1){
 \makebox(0,0)[t]{\begin{turn}{-90}\epsfig{figure=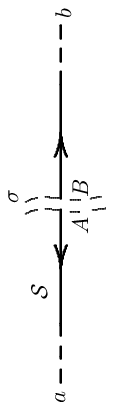,width=3.0cm}\end{turn}
}}
\caption{Точечный разрыв $\sigma$ на (локально плоском) 3-многообразии  
$\mathcal{S}$ с границами, обозначенными как $AB$ и $ab$.
Входящий поток (слева) и выходящий поток (справа) пробных частиц 
рассматривается как заряд частицы $\sigma$.} 
\label{fig:blackhole}
\end{figure}

Рассмотрим вращающееся 3-многообразие, например, сферу
 $\mathcal{S} \in \{\mathbb{S}^3\}$ 
радиуса $R_\mathcal{S} \in (-\infty, +\infty)$, образованную
безмассовой эластичной жидкостью  -- совокупностью мировых линий
пробных частиц. Вращение $\mathcal{S}$ соответствует 4-скорости  
$\mathbf{u}$ жидкости во всех возможных направлениях в каждой
точке пространственного сечения многообразия (см., например,
обсуждение вращающихся 3-многообразий в \cite{artin26}).
Ограничение на кривизну многообразия $\mathcal{S}$ 
возникает естественным образом из-за невозможности 
мгновенного изменения направления скорости пробных 
частиц, формирующих многообразие.
Пусть многобразие имеет выколотую точку, $\mathcal{S} \rightarrow \mathcal{S}\backslash\{0\}$, т.е. содержит точечный разрыв 
$\sigma$, моделирующий сингулярность  
(Рис.\ref{fig:blackhole}), который мы будем рассматривать в 
качестве базовой (простейшей) частицы, не имеющей никаких свойств, 
за исключением свойства обладания зарядом (входящий или выходящий 
поток пробных частиц, как показано на Рис.\ref{fig:blackhole}).
%

\begin{figure}[htb]
\begin{turn}{-90}
\centering 
\epsfysize=2cm
\includegraphics[scale=0.8]{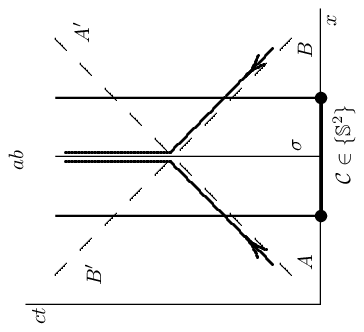}
\end{turn}
\caption{Мировые линии частиц потока $AA'$, $BB'$ (при отсутствии сингулярности) и 
$Aa$, $Bb$ (при входе в сингулярность $\sigma$, охваченную контрольной 
2-сферой $\mathcal{C}$). Во втором случае поток не сохраняется.}
 \label{fig:conservation}
\end{figure}

Если на $\mathcal{S}$ задана пробная поверхность $\mathcal{C}$
($\mathcal{C} \in \mathbb{S}^2$), не охватывающая сингулярностей, 
то входящий поток через эту поверхность должет быть равен 
выходящему потоку (Рис.\ref{fig:conservation}), что является обычным
условием сохранения потока. В случае, если поверхность $\mathcal{C}$
охватывает сингулярность $\sigma$, то исчезает либо входящий, 
либо выходящий поток, что противоречит принципу сохранения
энергии.
Для устранения противоречия поток должен быть направлен обратно
в $\mathcal{S}$, что можно сделать путем отождествления границ 
$ab$ с $AB$:
\begin{equation}
\begin{matrix}
{\rm a~\rightarrow~A} \\
{\rm b~\rightarrow~B} 
\end{matrix}
\label{eq:boundaries1}
\end{equation}
или путем их перекрестного отождествления:
\begin{equation}
\begin{matrix}
{\rm a~~~A} \\
 \hspace{-1.4mm} {\vspace{-0.9mm} \hspace{0.3mm}\nearrow 
 \hspace{-4.2mm} {\vspace{1.4mm} \searrow}}  \\
{\rm b~~~B} 
\end{matrix}
\label{eq:boundaries2}
\end{equation}
\noindent
Полученная форма является гипертором $\mathbb{T}^3$ (в первом случае), 
или бутылкой Клейна $\mathbb{K}^3$ (в случае с перекрестным отождествлением 
границ).
Это восстанавливает целостность потока и заменяет сингулярность
в точке $\sigma$ на топологическую особенность, представляющую
собой центральное отверстие (\glqqгорловину\grqq) гипертора 
или бутылки Клейна.  
Входящий (выходящий) поток $Q$ через данный объект по-прежнему
теряется, но полный поток в системе сохраняется.  
Как мы уже отметили в \S\,1, такая топологическая особенность 
соответствует несингулярному решению уравнений Эйнштейна, описывающему 
мост Эйнштейна-Розена (кротовую нору), свойства которого исследовались
многими авторами \cite{morris88}. Стоит обратить внимание на 
аналитическое решение, полученное недавно Шатским, Новиковым и Кардашевым 
\cite{shatskii08}, которое описывает бесконечное число сферических 
вселенных, соединенных между собой через кротовые норы. Авторы 
интерпретируют данную структуру как множество миров, возникающих
из квантового вакуума в разных областях пространства-времени 
и существующих параллельно и независимо друг от друга. 
С точки зрения модели хромоэлектрических преонов,
рассматриваемой здесь, данное решение соответствует
одной вселенной, содержащей бесконечное число динамически 
взаимодействующих топологических особенностей (простейших 
частиц). 

\section{Поле простейшей частицы}
\label{field}

\begin{figure}[htb]
\begin{turn}{-90}
\centering 
\epsfysize=2cm
\includegraphics[scale=0.8]{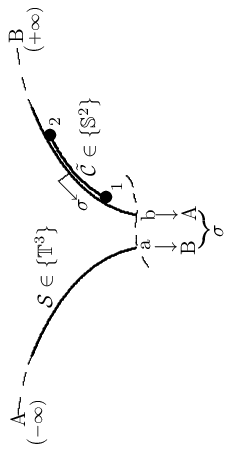}
\end{turn}
\caption{\glqqГорловина\grqq (топологическия особенность) $\sigma$ гипертора 
$\mathbb{T}^3$ или бутылки Клейна $\mathbb{K}^3$ и пробная сфера
$\tilde{\mathcal{C}} \in \mathbb{S}^2$, помещенная в $\sigma$ путем 
сдвига $\tilde{\mathcal{C}}$ по 
$\mathcal{S}$ в направлении топологической особенности.
}
\label{fig:twochoices}
\end{figure}

\begin{figure}[htb]
\begin{turn}{-90}\epsfig{figure=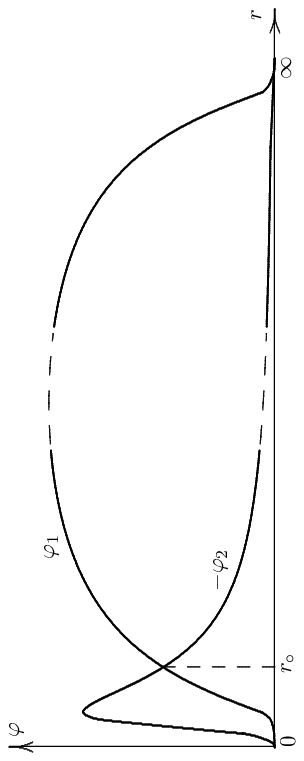, 
 width=3.5cm}\end{turn}
\caption{Два компонента равновесного поля, $\varphi_1$ и $\varphi_2$, 
соответствующие внутренней и внешней точкам пробной сферы 
${\Tilde{\mathcal{C}}}$.
}
 \label{fig:modifiedField}
\end{figure}

Рассматривая вышеописанные топологические особенности многообразия
в качестве простейших частиц мы должны учесть  
взаимодействие частиц между собой, 
что в нашем случае реализуется за счет давления и натяжения жидкости,
образующей многообразие $\mathcal{S}$. 
Соответствующее силовое поле, $\varphi$, можно определить 
через плотность потока, пересекающего произвольную двумерную поверхность  
$\mathcal{C}$, охватывающую $\sigma$. Поскольку входящий\,/\,выходящий 
поток пробных частиц изотропен, то поле является сферически-симметричным
и его величина приблизительно обратно пропорциональна квадрату 
расстояния между $\mathcal{C}$ и $\sigma$.
Однако охватить центральное отверстие ($\sigma$) 3-тора произвольной 
двумерной поверхностью топологически невозможно. 
Поэтому определение $\varphi$ в нашем случае следует изменить.
Пусть на $\mathcal{S} \in \{\mathbb{T}^3\}$  или $\{\mathbb{K}^3\}$
определена двумерная сфера $\Tilde{\mathcal{C}}$ радиуса  
$R_{\Tilde{\mathcal{C}}} \in (0,\infty)$ с центром в  
$\sigma$ (но не охватывая $\sigma$) -- как показано на Рис.\ref{fig:twochoices}; 
а именно, сдвигая $\Tilde{\mathcal{C}}$ вдоль $\mathcal{S}$ 
по направлению к  $\sigma$. 
Полный поток $\Tilde{Q}$ через $\Tilde{\mathcal{C}}$ равен нулю, поскольку
$\sigma$ не охвачена сферой $\Tilde{\mathcal{C}}$. 
Тем не менее, $\sigma$ делит $\Tilde{\mathcal{C}}$ на два полушария,
так что поток $\Tilde{Q}$ делится на две части, соответствующие 
внешнему и внутреннему полушариям сферы $\Tilde{\mathcal{C}}$.
Вычисляя плотность потока во внутренней  ``1'' и внешней  
``2'' точках сферы $\Tilde{\mathcal{C}}$ (Рис.\ref{fig:twochoices}),
можно разложить поле $\varphi$ на две составляющие:
\begin{equation}
\varphi(r)=\varphi_1(r) + \varphi_2(r),
\label{eq:superpos}
\end{equation}
где $r$ есть радиальное расстояние в сферических координатах с центром
в $\sigma$. Хотя плотность линий потока увеличивается при уменьшении 
$R_{\Tilde{\mathcal{C}}}$, она будет минимальна в начале отсчета (или даже равна
нулю в случае некомпактного многообразия), поскольку граница окрестности
начала отсчета $\sigma$ отождествлена с внешним (б\'ольшим) периметром тора.
Соответствующим граничным условием (для некомпактного многообразия)
будет  
\begin{equation}
\varphi_1(0)=\varphi_2(0)=0.
\label{eq:boundary1}
\end{equation}
Отсюда следует, что на некотором расстоянии от  
$\sigma$ плотность линий потока должна иметь максимум,
который предположительно соответствует верхнему пределу 
кривизны многообразия.
Изменяя радиус $R_{\Tilde{\mathcal{C}}}$ (например, от нуля до максимального
значения) точки ``1'' и ``2'' сферы ${\Tilde{\mathcal{C}}}$ перемещаются 
по многообразию в противоположных направлениях:
\begin{equation}
\begin{matrix}
\text{``1''}: ~~~A \rightarrow a \\
\text{``2''}: ~~~a \rightarrow A 
\end{matrix}
\label{eq:paths}
\end{equation}
что приводит к асимметрии между двумя компонентами поля. 
А именно, $\varphi_2(r)$ будет расти от минимума (нуля) до некоторого 
максимального значения, а затем опять спадет до минимума 
в конце пути точки ``2'' (Рис.\,\ref{fig:modifiedField}). 
Компонент $\varphi_1(r)$ будет в основном расти вдоль большей 
части траектории точки  ``1''. Но, конечно, в конце этой траектории
 $\varphi_1$ спадет до минимума (нуля).
Таким образом, вторым граничным условием для некомпактного многообразия
будет  
\begin{equation}
\varphi_1(\infty)=\varphi_2(\infty)=0.
\label{eq:boundary2}
\end{equation}

Здесь однако следует принять во внимание кручение, 
которое в случае 3-многообразия имеет три степени свободы 
\cite{ritis83},
что приводит к нелинейному уравнению Иваненко-Гейзенберга 
\cite{ivanenko38} и неабелевым степеням свободы. 
Соответствующее поле будет иметь топологическое квантовое
число -- цветной аналог спиральности в гидродинамике \cite{jackiw00}.
Такой характер взаимодействия соответствует притяжению
и отталкиванию между цветовыми зарядами, используемыми 
в хромодинамике \cite{suisso02}: две одноименно-заряженных частицы с разноименными 
цветами притягиваются, в противном случае они отталкиваются.
Приближенная антисимметрия между $\varphi_1$ и $\varphi_2$ в окрестности
начала отсчета подразумевает существование расстояния 
равновесия   
$r_\circ>0$ такого, что
\begin{equation}
\varphi_1(r_\circ)=-\varphi_2(r_\circ),
\label{eq:equilibrium2}
\end{equation}
так что компоненты поля полностью компенсируют друг друга.
Это нарушает начальную сферическую симметрию поля, 
так как, например, в простейшем случае задачи двух тел основное 
состояние является диполем, ориентация которого в пространстве
определяет преимущественное направление. Симметрия масштабной 
инвариантности здесь также нарушается, поскольку расстояние 
равновесия $r_\circ$ фиксирует преимущественную единицу
измерения расстояний для всех композитных систем, состоящих 
из цветных преонов. 

\section{Простейшие системы цветных преонов}
\label{hierarchy}

Рассмотрим некоторые самые простые структуры, основанные на 
полях (\ref{eq:superpos}) с граничными условиями (\ref{eq:boundary1}) и
 (\ref{eq:boundary2}) и условием равновесия (\ref{eq:equilibrium2}).
Следующая форма для компонентов этого поля:
\begin{equation}
\begin{matrix}
\varphi_1(r)=&\varkappa\exp(-\kappa r^{-1}) \\
\varphi_2(r)=&{-\varphi'_1(r)~~~~~~~~~~~~~~~}
\end{matrix}
\label{eq:exp1}
\end{equation}
удовлетворяет условиям (\ref{eq:boundary1}) и (\ref{eq:equilibrium2}).
Она не удовлетворяет второму граничному условию (\ref{eq:boundary2}),
но для случаев с типичными расстояниями между частицами
порядка $r_\circ$ это не существенно,
поэтому на близких расстояниях мы можем принять форму 
(\ref{eq:exp1}) в качестве простого примера поля, иллюстрирующего
работоспособность предлагаемой модели. Этому полю соответствует
потенциал: 
\begin{equation}
V(r)=(1-r)\exp(-\kappa r^{-1})-{\rm Ei}(-\kappa r^{-1}).
\label{eq:spotential}
\end{equation}
Для простоты (а также для того, чтобы избежать использования
свободных параметров в нашей модели) масштабный коэффициент 
$\kappa$ можно принять равным единице. Коэффициент 
$\varkappa=\pm1$ в (\ref{eq:exp1}) обозначает полярность поля 
и выбирается таким образом, чтобы воспроизвести вышеупомянутое свойство 
притяжения\,/\,отталкивания  между частицами
с разными цветовыми зарядами.

В многокомпонентных системах поле такого рода приводит 
к потенциальной поверхности с множественными
локальными минимумами, что ведет к кинематическим ограничениям 
топологического свойства. 
Это определяет единственность набора возможных
конфигураций преонов, простейшими из которых являются диполи и триполи,
сформированные, соответственно, из двух и трех хромоэлектрических преонов.
Очевидно, что поля триполей -- бесцветны, но их 
полная бесцветность достигается лишь на бесконечности. 
На близких расстояниях поля будут хроматически
поляризованы (так как центры частиц разнесены на некоторое расстояние), 
что позволяет триполям взаимодействовать друг с другом и 
образовывать более сложные структуры (в основном, путем
соединения в цепочки полюсами друг к другу).
\begin{figure}[htb]
\begin{turn}{-90}\epsfig{figure=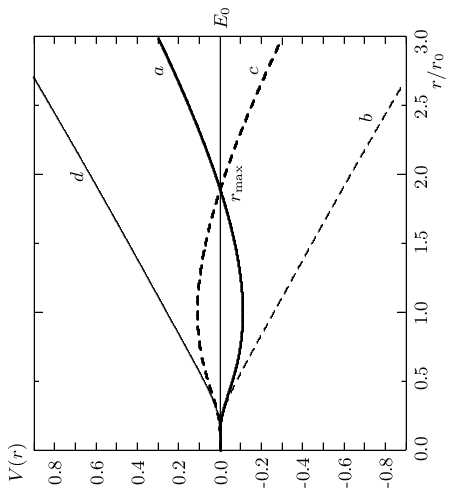, 
 width=7.0cm}\end{turn}
\caption{Равновесный потенциал $V(r)$ соответствующий
системе цветного диполя;\,\, ($a$)\,две одноименно-заряженные
частицы с разноименными цветами; 
($b$) то же, но для частиц с одноименными цветами; 
($c$)\, разноименно-заряженные частицы с одноименными цветами; 
и ($d$) разноименно-заряженные частицы с разноименными цветами.
 \label{fig:spotential}
}
\end{figure}

Потенциал (\ref{eq:spotential}) графически показан 
на Рис.\,\ref{fig:spotential}\,$a$.  
Это типичный двухъямный потенциал, приводящий 
к хаотическим колебаниям и стохастическим резонансам 
\cite{wiggins89}. В космологии он ведет к образованию доменных 
перегородок в периоды фазовых переходов на ранних этапах 
расширения вселенной. Известно также, что этот потенциал является
самокалиброванным, так как в нем определены единицы 
длины, времени и энергии через, соответственно, 
расстояние между потенциальными ямами, частоту колебаний
в инвертированном потенциальном барьере и высоту барьера.  
Мы будем использовать половину расстояния между потенциальными
ямами (расстояние равновесия $r_\circ$, соответствующее
минимуму потенциала) в качестве единицы длины для данной модели. 
Единица скорости, $v_\circ$, тоже является внутренней характеристикой
потенциала (\ref{eq:spotential}) в том смысле, что, если задана 
энергия $E_0$, то можно вычислить соответствующую скорость частицы
в этом потенциале. Например, в системе из двух преонов 
с потенциалом (\ref{eq:spotential}) скорость вычисляется как   
\begin{equation}
v(r)=\sqrt{2(E_0-V(r))/\hat{m}},
\label{eq:speed2particles}
\end{equation}
где $\hat{m}^{-1}=m_1^{-1}+m_2^{-1}$
есть приведенная масса (здесь $\hat{m}=\frac{1}{2}$, поскольку 
мы используем $m_1=m_2=1$). 
Задав некоторое значение энергии $E_0$, например приняв его
равным нулю в точке $r=r_{\rm max}$, можно вычислить максимальную
скорость  $v_{\rm max}=v(r_\circ)\approx 0.937v_\circ$, 
выраженную в единицах, определяемых масштабными коэффициентами
функции (\ref{eq:exp1}). При этом определяется также масштаб
шкалы времени в виде единичного временного интервала $t_\circ$, 
такого, что $v_\circ t_\circ = r_\circ$, а также и другие 
необходимые физические единицы, как, например, единица углового
момента $L_\circ$, которая соответствует преону единичной массы 
$m_\circ$, движущемуся с единичной скоростью $v_\circ$ по
круговой орбите единичного радиуса $r_\circ$.
Единицу массы $m_\circ$ можно определить, вычисляя энергию поля 
$\varphi$ для преона:
\begin{equation}
m^2= \frac{1}{2\pi}\,\int\limits_{\mathcal{S}} \varphi^2 \,d\mathcal{S} \,=\,1 
\label{eq:masssquared}
\end{equation}
без учета первого компонента поля, $\varphi_1$,
поскольку интеграл  
$\int \varphi_1^2 \,d\mathcal{S}$ расходится
и очевидно не может быть использован для этой цели.
Если временно отбросить третью цветовую полярность
и рассмотреть систему  
$d^+_{\rm y}=\sigma^+_{\rm r} \, \sigma^+_{\rm g}$
-- заряженный цветной диполь с энергией $E_0=0$, то можно заметить,
что (колебательное) движение зарядов в этой системе ограничено 
областью  
$(0,r_{\rm max})$, где $r_{\rm max}\simeq 1.894r_\circ$ 
(см. Рис.\,\ref{fig:spotential}).  
Индекс {\scriptsize ``y''} (yellow=$\overline{\rm blue}$) 
обозначает, что данная структура
неполна в отношении полярности поля, 
обозначенной синим цветом ($\varphi_1^{\rm b}$), и что заряженный цветной 
диполь $d^+_{\rm y}$ на самом деле не может существовать 
в свободном состоянии, так как его энергия бесконечна 
(если конечно мы не пренебрегаем третьим цветом).
Суммарный заряд диполя $d^+_{\rm y}$ равен $+2q_\circ$. 
Его масса $m_{d^+_{\rm y}}$ в первом приближении равна 
$2m_\circ$, что является точным значением, когда центры 
преонов $\sigma^+_{\rm r}$ и $\sigma^+_{\rm g}$ совпадают 
и компоненты $\varphi_1^{\rm r}$ и $\varphi_1^{\rm g}$ 
поля полностью взаимно уничтожаются (компонентом
$\varphi_1^{\rm b}$ пренебрегаем).

Если компоненты системы $d^+_{\rm y}$ разнесены на некоторое расстояние
$D$, то поля $\varphi_1^{\rm r}$ и $\varphi_1^{\rm g}$ компенсируют друг 
друга только частично, и масса системы превысит $2m_\circ$ на величину,
которую мы в дальнейшем будем называть эксцессом массы и которая 
может быть вычислена по формуле (\ref{eq:masssquared}). 
В основном состоянии ($D=r_\circ$) масса этой системы 
составляет $\approx 2.04\,m_\circ$.

В случае системы из двух преонов с одинаковыми цветами,
например, $\sigma^+_{\rm r} \, \sigma^+_{\rm r}$, 
следует инвертировать знак $\varkappa$ в (\ref{eq:exp1}),
что приведет к отталкивающему потенциалу (кривая $b$ на 
Рис.\,\ref{fig:spotential}) с максимумом в начале координат.
Это означает, что система  
$\sigma^+_{\rm r}\sigma^+_{\rm r}$ 
не может быть сформирована в принципе, даже если пренебречь
третьей полярностью.   

Минимумы потенциалов, соответствующих парам преонов с противоположными
электрическими зарядами (кривые $c$ и $d$ на 
Рис.\,\ref{fig:spotential}) свидетельствуют о том, что 
нейтральные цветные диполи  
$d^0_{\rm y}=\sigma^+_{\rm r} \, \sigma^-_{\rm g} \hspace{0.05cm}$
и
$d^0_{\rm r}=\sigma^+_{\rm r} \, \sigma^-_{\rm r}$
в принципе могут быть сформированы (конечно, если опять 
пренебречь третьей цветовой полярностью). 
Тогда при $D=0$ система $d^0_{\rm y}$ будет 
иметь исчезающе-малую массу 
(100\% дефект массы), поскольку в этом случае 
противоположные знаки будут иметь не только поля 
$\varphi_1$, но и поля $\varphi_2$, 
так что все поля компонентов этой системы будут взаимно
скомпенсированы. При $D>0$ масса диполя $d^0_{\rm y}$ будет 
расти почти линейно с расстоянием.

Бесцветное связанное состояние трех одноименно-заряженных преонов
с взаимно-дополняющими цветами (триполь)  
$Y^+=\sigma^+_{\rm r} \, \sigma^+_{\rm g} \, \sigma^+_{\rm b} \hspace{0.05cm}$
или
$Y^-=\sigma^-_{\rm r} \, \sigma^-_{\rm g} \, \sigma^-_{\rm b}\,,$
имеет конечную массу. Если начальная энергия $E_0=0^{\text{-}}$, то 
радиус триполя $R_Y$ будет осциллировать в пределах 
от нуля до $R_Y^{\rm max} \simeq 1.09r_\circ$.
Система триполя является стабильной, так как для того, чтобы 
удалить любой из
ее компонентов на бесконечное расстояние, требуется бесконечная 
энергия. В основном состоянии компоненты триполя разнесены на расстояние 
$D=r_\circ$, соответствующее равновесному радиусу 
$R_Y=r_\circ/\sqrt{3}$, который минимизирует потенциал системы.
Поскольку центры компонентов триполя не совпадают, их поля 
$\varphi_1$ компенсируют друг друга лишь частично, так что 
масса основного состояния триполя будет иметь эксцесс около  
$0.199\,m_\circ$ на каждый преон, по сравнению с массой при
$R_Y=0$.

\section{Двух- и трехкомпонентные цепочки триполей}

\begin{figure}[htb]
\hspace{-1.5cm}
\begin{turn}{-90}\epsfig{figure=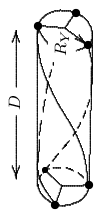, 
 width=3.5cm}\end{turn}
\put(50,10){
 \makebox(0,0)[t]{\epsfig{figure=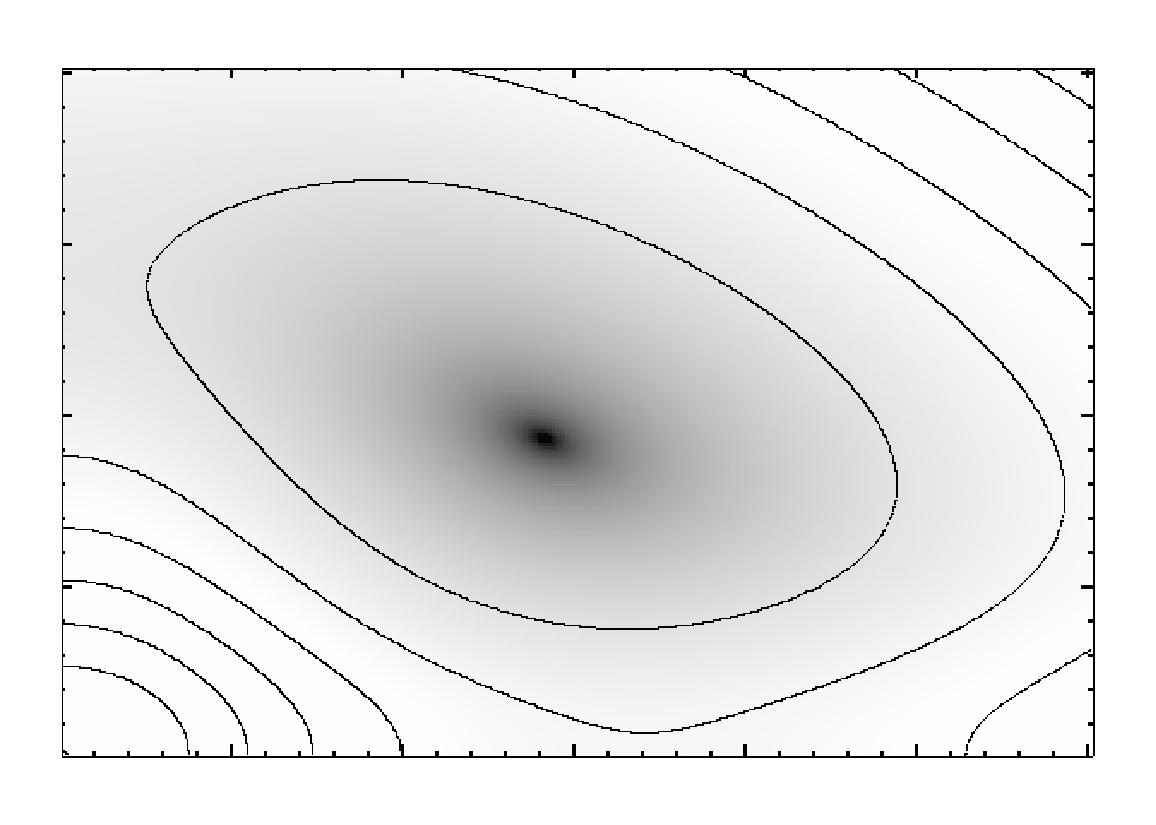,width=5.5cm}
 \put(-159,100){
 \makebox(0,0)[t]{\scriptsize{$R_Y$}}}
 \put(-10,5){
 \makebox(0,0)[t]{\scriptsize{$D$}}}
 \put(-152,5){
 \makebox(0,0)[t]{\footnotesize{$0$}}}
 \put(-105,5){
 \makebox(0,0)[t]{\footnotesize{$1.0$}}}
 \put(-58,5){
 \makebox(0,0)[t]{\footnotesize{$2.0$}}}
 \put(-160,34){
 \makebox(0,0)[t]{\footnotesize{$0.5$}}}
 \put(-160,59){
 \makebox(0,0)[t]{\footnotesize{$1.0$}}}
 \put(-160,82){
 \makebox(0,0)[t]{\footnotesize{$1.5$}}}
}}
\put(220,10){
 \makebox(0,0)[t]{\epsfig{figure=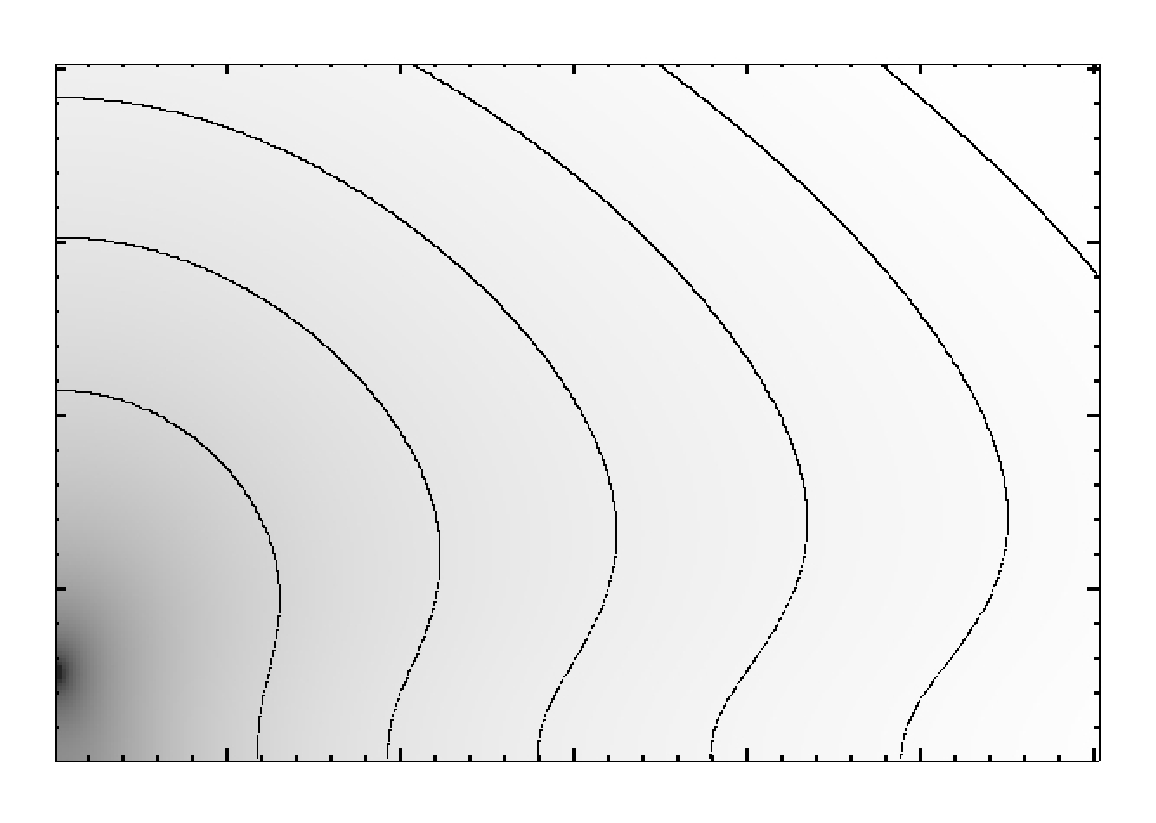,width=5.5cm}
 \put(-161,100){
 \makebox(0,0)[t]{\scriptsize{$R_Y$}}}
 \put(-10,5){
 \makebox(0,0)[t]{\scriptsize{$D$}}}
 \put(-152,5){
 \makebox(0,0)[t]{\footnotesize{$0$}}}
 \put(-105,5){
 \makebox(0,0)[t]{\footnotesize{$1.0$}}}
 \put(-58,5){
 \makebox(0,0)[t]{\footnotesize{$2.0$}}}
 \put(-160,34){
 \makebox(0,0)[t]{\footnotesize{$0.5$}}}
 \put(-160,59){
 \makebox(0,0)[t]{\footnotesize{$1.0$}}}
 \put(-160,82){
 \makebox(0,0)[t]{\footnotesize{$1.5$}}}
}}
 \put(45,9){
 \makebox(0,0)[t]{a}}
 \put(220,10){
 \makebox(0,0)[t]{b}}
\caption{Потенциальная энергия цепочки, сформированной из двух триполей
$Y$, разнесенных на расстояние $D$ и имеющих радиусы $R_Y$ 
(схема слева); (a)\, в случае двух одноименно-заряженных триполей; 
(b)\, в случае двух противоположно-заряженных триполей.  
}
\label{fig:doubleTripole}
\end{figure}

Вследствие хроматический поляризации полей триполя, различные 
триполи могут взаимодействовать друг с другом и формировать 
новые связанные состояния. 
Простейшим из них является твухкомпонентная цепочка, образованная
либо из двух одноименно-заряженных:  
$\gamma^+=Y^+ \, Y^+$
или
$\gamma^-=Y^- \, Y^-\,,$
либо из противоположно заряженных триполей: 
$\gamma^\circ=Y^+ \, Y^-\,.$
Очевидно, что триполи в этих системах будут ориентированы 
полюсами друг к другу, с разворотом на $180^\circ$ (см. схему
слева на Рис.\,\ref{fig:doubleTripole}). 
Потенциальные энергии для пар одноименно- и
разноименно-заряженных триполей как функции радиусов 
триполей $R_Y$, и расстояния $D$ между триполями показаны 
на Рис.\,\ref{fig:doubleTripole}\,(a) и (b), соответственно.  

Минимум потенциала для второй (электрически нейтральной) системы 
$\gamma^\circ$, Рис.\,\ref{fig:doubleTripole}\,(b), 
почти совпадает с началом координат. Это означает, что как
электрические, так и хроматические поля в этой системе
взаимно компенсируются практически полностью. 
Поэтому в основном состоянии эта система будет иметь 
пренебрежимо малую массу и не будет взаимодействовать 
с другими подобными (нейтральными) системами. 
Так что в каждой точке пространства можно поместить 
сколь угодно много таких систем (плотность их распределения
к сожалению нельзя вывести из первых принципов, использованных
в данной модели). Имея нулевую массу, эти частицы 
будут двигаться с максимально возможной скоростью во всех
возможных направлениях, что можно рассматривать как 
идеальный газ нейтральных частиц.
Тем не менее, в окрестности электрического или цветового
заряда компоненты системы $\gamma^\circ$ будут поляризованы
либо электрически, либо хроматически, что превратит данную
систему в элетрический или цветной диполь.
Это свойство частиц $\gamma^\circ$ является важным, 
поскольку в рамках данной модели можно использовать 
газ частиц $\gamma^\circ$ для моделирования поляризуемой 
среды (вакуума), в которой происходят движения и 
взаимодействия всех остальных частиц. Диполь-дипольные 
взаимодействия между поляризованными частицами $\gamma^\circ$
будут приводить к формированию мгновенных преонных конфигураций
различной сложности, которые можно рассматривать в качестве 
источника новых частиц при условии, что энергия, приводящая 
к поляризации, достаточна для полного разделения компонент 
$\gamma^\circ$ (в противном случае эти конфигурации можно 
использовать для моделирования виртуальных частиц).
Следует заметить, что частица $\gamma^\circ$ является единственной
формой материи, которая может существовать в самой начальной фазе 
расширения вселенной, когда доступный объем еще слишком мал для 
формирования б\'ольших структур (то есть когда радиус вселенной
не превышает $\approx 0.5 r_\circ$). Движения этих частиц будут
стохастичными из-за наличия нестабильной стационарной точке 
потенциала в начале отсчета. 

\begin{figure}[htb]
\begin{turn}{-90}\epsfig{figure=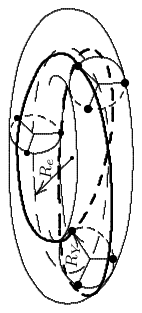, 
 width=3.5cm}\end{turn}
\put(100,10){
 \makebox(0,0)[t]{\epsfig{figure=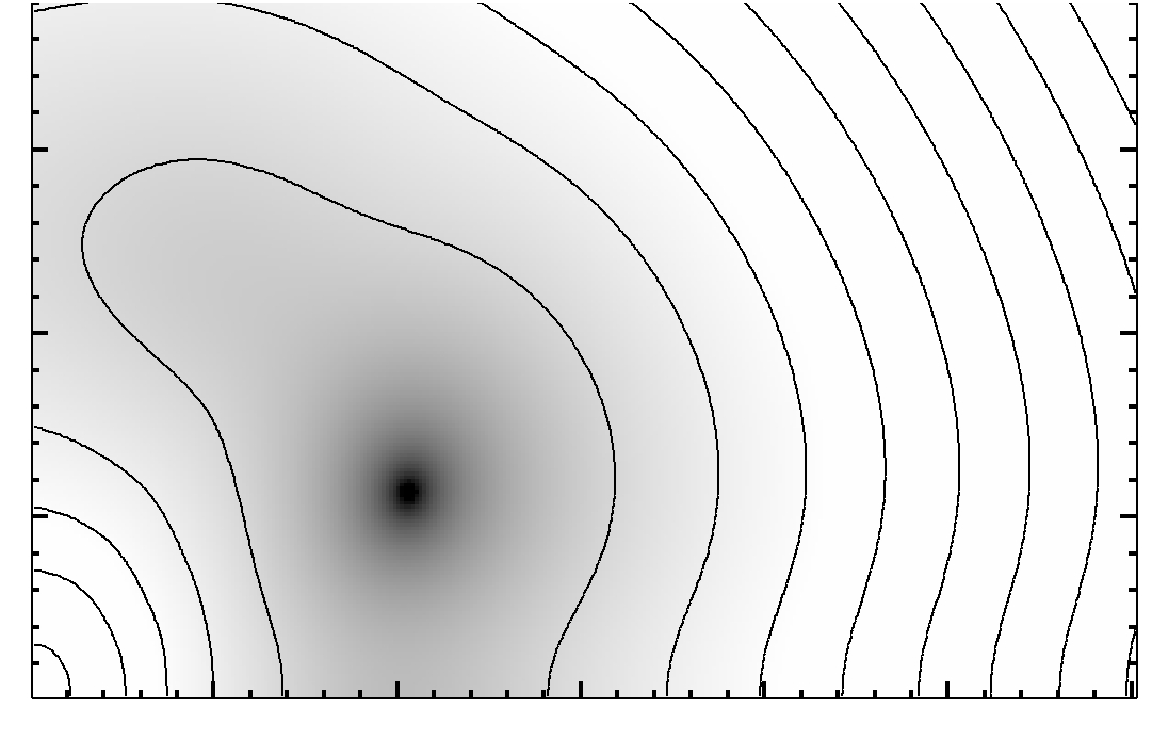,width=5.5cm}
 \put(-163,100){
 \makebox(0,0)[t]{\scriptsize{$R_Y$}}}
 \put(-3,4){
 \makebox(0,0)[t]{\scriptsize{$R_e$}}}
 \put(-155,2){
 \makebox(0,0)[t]{\footnotesize{$0$}}}
 \put(-105,2){
 \makebox(0,0)[t]{\footnotesize{$1.0$}}}
 \put(-56,2){
 \makebox(0,0)[t]{\footnotesize{$2.0$}}}
 \put(-164,34){
 \makebox(0,0)[t]{\footnotesize{$0.5$}}}
 \put(-164,59){
 \makebox(0,0)[t]{\footnotesize{$1.0$}}}
 \put(-164,82){
 \makebox(0,0)[t]{\footnotesize{$1.5$}}}
}}
\caption{Потенциальная поверхность, соответствующая 
цепочке $e^\pm$ радиуса $R_e$, сформированной из трех одноименно
заряженных триполей $Y$, имеющих радиусы $R_Y$ 
(схема слева). Радиусы варажены в единицах $r_\circ$. 
}
\label{fig:minimum}
\end{figure}

По аналогии с трехкомпонентной системой преонов $Y^\pm$
мы можем рассматривать трехкомпонентную цепочку, $e^\pm$, 
сформированную из одноименно-заряженных триполей:
$e^+=Y^+ \, Y^+ \, Y^+ \hspace{0.05cm}$
или
$e^-=Y^- \, Y^- \, Y^-\,.$
Потенциальная энергия этой цепочки минимальна, когда цепочка 
замкнута в кольцо, а ее компоненты (триполи) повернуты 
друг относительно друга на $120^\circ$
(см. схему слева на Рис.\,\ref{fig:minimum}). 
Минимум потенциальной поверхности, показанной на Рис.\,\ref{fig:minimum}, 
соответствует статической конфигурации равновестия цепочки. 

\noindent
Однако на самом деле такое статическое состояние будет нестабильным,
потому что триполи в цепочке сохраняют вращательную и поступательную
степени свободы (вокруг и вдоль их общей кольцевой оси) и будут
двигаться под действием других существующих частиц, например,
под стохастическим воздействием вышеупомянутого газа нейтральных
двухкомпонентных цепочек триполей. 
Динамические параметры данной системы будут значительно 
отличаться от статического случая, показанного на 
Рис.\,\ref{fig:minimum}, так как каждый из трех вращающихся 
триполей в этой системе будет генерировать магнитное поле,
модифицирующее (и стабилизирующее) движения двух других
триполей \cite{pati81}.  
За счет стандартного механизма динамо эти движения сформируют
тороидальные и полоидальные магнитные поля, которые, в свою
очередь, будут поддерживать вращательное и орбитальное 
движения триполей в кольце. 

Цветные токи, соответствующие движению каждого их преонов
данной системы, закручены в виде винтовой линии 
(кривой Смэйла-Вильямса), которая при замыкании делает 
поворот на $\pi$ радиан вокруг центральной кольцевой 
оси системы -- либо по часовой стрелке, либо против часовой
стрелки (в системе, на Рис.\ref{fig:minimum} показаны токи, 
закрученные против часовой стрелки).
Такой сдвиг (дислокация) фазы является инвариантным свойством
и называется топологическим зарядом \cite{kleman83} или индексом 
дислокации, знак которого соответствует направлению закрутки 
(по- или против часовой стрелки), а величина определяется 
числом оборотов, приходящихся на один орбитальный период.
В этих терминах, сдвиг по фазе токов в структуре $e^\pm$ 
на $\pi$ радиан соответствует топологическому заряду 
$S=\pm \frac{1}{2}$, который можно идентифицировать как 
внутренний угловой момент (спин) частицы. 

Траектория каждого преона, входящего 
в один из триполей, в точности совпадает с траекториями
двух других преонов, принадлежащих двум соседним триполям 
структуры и имеющим цветовые заряды, дополнительные к первому
преону. Таким образом получается, что токи, образующие структуру $e^\pm$,
динамически бесцветны и, усредненное по времени, поле этой
частицы может иметь только две полярности, соответствующие 
полярностям стандартного электрического поля.

Можно заметить, что, кроме простейших связанных состояний 
хромоэлектрических преонов, описанных в предыдущих разделах,
поле (\ref{eq:exp1}) генерирует множество других
структур. Например, цепочка из пар триполь-антитриполь 
обладает симметрией, подобной той, что имеет структура $e^\pm$, 
что также приводит к замыканию цепочки 
в кольцо $\nu_e$ с минимальным числом компонентов, равным шести
парам (двенадцати триполям). 

По своим свойствам, которые включают спин, заряд, 
гиромагнитное оношение и четность \cite{yershov07,yershov05}, 
кольцевые структуры, состоящие из трех и двенадцати триполей,
($e^\pm$ и $\nu_e$) можно идентифицировать с электроном и электронным
нейтрино, соответственно.  
Различные комбинации этих структур
(включая также собственно триполь $Y^\pm$) составляют
иерархию структур, которую можно идентифицировать с наблюдаемым 
в природе разнообразием элементарных частиц.
Число элементов в каждой структуре однозначно определяется 
минимумом потенциальной энергии этой структуры, так что 
данная иерархия является единственно возможной.

\section{Космологическая сингулярность}
\label{cosmologicalSingularity}

Преонные структуры, обсуждавшиеся в предыдущем параграфе,
соответствуют многообразию  $\mathcal{S}$ большого размера, 
но при уменьшении доступного объема ситуация должна измениться. 
Предположим, что наше многообразие $\mathcal{S}$ эволюционирует
(расширяется или сжимается), а его кривизна изменяется во времени.
Во время фазы сжатия растущая кривизна достигнет верхнего
предела, соответствующего внутренней кривизне простейших частиц.
Поскольку потенциал (\ref{eq:spotential}) имеет стационарную
точку в начале отсчета, то все простейшие частицы на многообразии 
$\mathcal{S}$ будут сжаты в минимальный объем, с их центрами
совпадающими в начале отсчета. На этом этапе скорость сжатия 
и кривизна многообразия -- максимальны, а объем -- минимален,
но не равен нулю.   
Такая бесструктурная суперпозиция простейших 
частиц (отсутствие материи) соответствует фазе де Ситтера, и, 
как известно, является приемлемым кандидатом на роль модели 
несингулярного начального состояния вселенной 
\cite{gliner75,markov83}. 
 
Помимо проблемы космологической сингулярности, существует еще 
целый ряд важных космологических проблем, которые должны
быть приняты во внимание в любой модели вселенной. 
Здесь мы кратко очертим круг этих проблем и обозначим возможности
их решения в рамках модели с хромоэлектрическими полями. 

\vspace{0.2cm} 
{\it \underline{Проблема энтропии}. \,}
Проблема энтропии имеет отношение к \glqqспециальности\grqq \,начального
состояния вселенной  \cite{penrose82}.
Сам факт существования второго закона термодинамики подразумевает,
что начальное состояние вселенной должно было иметь очень
малое значение энтропии \cite{penrose89}.
Согласно этому закону, в циклических моделях вселенная
должна увеличиваться в размерах от цикла к циклу \cite{tolman31}. 
При обратной экстраполяции во времени это приводит к той же самой
проблеме начальной сингулярности, которую циклические модели 
собственно и стараются разрешить. Для того, чтобы этого избежать,
избыток энтропии должен каким-либо образом изыматься из вселенной. 
Например, в модели осциллирующей вселенной Брауна и Фрамптона 
\cite{brown07} это осуществляется за счет фантомной темной энергии
и механизма дефляции. Авторы этой модели предложили, что в момент каждого 
следующего поворота цикла осциллирующей вселенной от расширения к сжатию 
остается только один причинно-связанный участок многообразия.
Остальные участки сжимаются независимо друг от друга в виде отдельных
вселенных. При этом избыток энтропии полностью изымается 
из нашей вселенной в тот момент, когда от масштабного фактора
остается лишь ничтожная часть, поэтому вселенная начинает каждый 
последующий цикл с пренебрежимо малым значением энтропии. 
Обсуждаемая здесь модель содержит гораздо более радикальное решение
проблемы энтропии. 
А именно, как уже упоминалось, в момент наибольшего сжатия вселенной 
простейшие частицы теряют все свои степени свободы и их центры полностью 
совмещаются. Поэтому энтропия такого начального состояния 
минимизируется (фактически, стремится к нулю), что 
отличается от общепринятого суждения о том, 
что энтропия никогда не уменьшается
\cite{penrose79}. 
Но на самом деле, такой результат логичен и просто отражает тот факт,
что энтропия как макроскопический статистический параметр
не может использоваться для описания одноэлементной микроскопической
системы, какой является сколлапсировавшая вселенная, описываемая
с помощью полей (\ref{eq:exp1}).
Конечно, энтропия этой вселенной будет постоянно расти как во время
фазы расширения, так и сжатия. Но в конце фазы сжатия вселенная неизбежно 
войдет в вышеописанное упорядоченное одноэлементное состояние,
означающее, что после каждого цикла вселенная будет полностью обновлена.
Таким образом в нашей модели проблема постоянно растущей
энтропии, которая характерна для многих циклических моделей
вселенной, полностью преодолена. По той же самой причине здесь не возникает
также и проблема производства огромной энтропии на предыдущем цикле
вследствие образования черных дыр \cite{penrose90}. 
Впервые возможность подобного \glqqхолодного\grqq начального состояния 
вселенной была рассмотрена Зельдовичем и Новиковым \cite{zeldovich75}. 
Они расчитали сегодняшнее значение энтропии вселенной, которое соответствовало
бы такому начальному состоянию и нашли, что их результат на удивление
хорошо согласуется с наблюдаемым значением энтропии (около $10^9$ на барион). 
В случае де ситтеровского начального состояния аналогичное значение 
было впервые расчитано в работе \cite{gliner75}.

\vspace{0.2cm} 
{\it \underline{Проблема \glqqотскока\grqq}. \,}
В свое время Лифшиц и Халатников \cite{lifschitz63} обсуждали возможность
пересечения мировых линий пробных частиц в одной и той же точке,
но с проходом мимо друг друга -- в случае неоднородной и неизотропной 
модели коллапсирующей вселенной. В подобной модели частицы коллапсирующей 
вселенной продолжают движение в том же направлении, и вселенная 
предстает как-бы \glqqотскакивающей\grqq (расширяющейся) после коллапса. 
Однако перспективы этой модели были серьезно подорваны после 
появления теорем Пенроуза и Хокинга о сингулярности \cite{penrose69}.
В нашем случае естественный отскок Лифшица-Халатникова может
происходить при $\mathbb{K}^3$-топологии вселенной, так как данная
топология позволяет переход от фазы сжатия к фазе расширения с 
максимальной скоростью и без изменения направления движения 
пробных частиц многообразия (в этом случае многообразие
просто выворачивается наизнанку без изменения его топологии). 
В отличие от этого, гиперсфера $\mathbb{S}^3$ или гипертор $\mathbb{T}^3$ 
не могут гомеоморфно перейти от сжатия к расширению, что 
оставляет нам только одну возможность: топология вселенной должна быть
топологией $\mathbb{K}^3$.

\vspace{0.2cm} 
{\it \underline{Проблема плоскостности пространства}. \,}
Из наблюдений следует, что вселенная должна быть пространственно
плоской, что в современной космологии рассматривается как невероятное
совпадение, так как плоская вселенная является 
лишь частным случаем среди огромного числа других возможностей. 
В обсуждаемой модели данная проблема решается без использования
тонкой настройки или антропного принципа, которые обычно используются
в этом случае  
(см., например, \cite{tangherlini93,durrer96} или 
обзор современного состояния проблемы в \cite{lahav04}). 
Известно, что многообразие $\mathbb{K}^3$ принадлежит к классу 
евклидовых, и что оно является конечным и геодезически полным 
\cite{thurston84}. Это означает, что любая метрика на бутылке Клейна
конформно эквивалентна плоской метрике \cite{jakobson06}. Более того,
в нашем случае глобальная геометрия пространственного сечения 
этого многообразия должна быть близкой к евклидовой на любой 
стадии эволюции многообразия. Это видно из того простого факта,
что в случае с $\mathbb{K}^3$ многообразие в каждой своей точке 
вывернуто наизнанку, так что
кривизны \glqqвнутренней\grqq \, и \glqqвнешней\grqq \, областей 
равны друг другу по величине, но обратны по знаку. Наблюденная 
полная кривизна в таком случае будет почти равной нулю, за исключением
остаточного эффекта, обусловленного наличием локальных искривлений, 
соответствующих материи (простейшим частицам и их комбинациям). 
Таким образом, обсуждаемая модель подразумевает
именно плоскую вселенную, какая и наблюдается в действительности 
\cite{spergel03}. 

\vspace{0.2cm} 
{\it \underline{Нарушение зарядовой симметрии}. \,}
Обсуждаемая модель является симметричной в отношении числа 
положительных и отрицательных зарядов (правая и левая 
части Рис.\ref{fig:blackhole}), поскольку каждая пара
противоположно-заряженных частиц рассматривается как 
проявление одной и той же топологической особенности.  
Нарушение зарядовой симметрии в данной модели можно объяснить,
рассматривая траектории пробных частиц входящего и выходящего потоков,
соответствующих отрицательному и положительному зарядам. 
Эти траектории различаются тем, что проходят по разным областям 
-- по \glqqвнутреннему\grqq и \glqqвнешнему\grqq \, сечениям 
многообразия $\mathbb{K}^3$. Различие в траекториях приводит 
к разным энергиям этих двух сечений, потому что, ввиду наличия верхнего 
предела кривизны, эти сечения в точности совместить друг с другом
невозможно, и они всегда будут иметь небольшую разницу в объемах
(за исключением мгновенной конфигурации, соответствующей 
фазе максимального сжатия). Знак разности между этими энергиями
меняется после каждого коллапса многообразия, так что, если
рассматривать полный цикл расширения-сжатия, то CPT-симметрия 
восстанавливается.

\vspace{0.2cm}  
{\it \underline{Проблема горизонта}. \,}
В ряде моделей проблема горизонта решается (наряду с некоторыми
другими космологическими проблемами) за счет принятия гипотезы
о переменности фундаментальной скорости взаимодействия 
\cite{albrecht99,chakraborty02}.
В рассматриваемой здесь модели величина фундаментальной скорости
предположительно связана со скоростью потока, формирующего 
вращающееся многообразие (см. \S\,2).
В расширяющейся вселенной с вращением количественная характеристика
вращения уменьшается \cite{goedel49} и, соответственно, 
уменьшается фундаментальная скорость, которая имеет максимальное 
значение в момент, когда вселенная переходит от фазы сжатия к 
расширению и минимальна когда вселенная достигает максимального
размера. 
В нашем случае переменность фундаментальной скорости следует из
необходимости сохранения момента импульса вращающегося
многообразия \cite{fan50}. Так что модель с хромоэлектрическими преонами
полностью принимает в себя уже достаточно хорошо проработанную 
теорию изменяющейся фундаментальной скорости, которая является
одной из альтернатив стандартной инфляционной модели
и которая ведет к предсказаниям пертурбационного спектра возмущений, 
однородности и плоскостности вселенной, сходным с получаемыми
в рамках инфляционной модели \cite{magueijo03}.  

\vspace{0.2cm} 
{\it \underline{Формирование структуры}. \,}
Потенциал (\ref{eq:spotential}) с нестабильной стационарной точкой 
в начале отсчета характерен тем, что в системах с этим потенциалом
симметрия спонтанно нарушается.   
Это приводит к разделению цветных зарядов на определенном этапе 
расширения многообразия и к самоорганизации этих зарядов  
в сложные структуры.  
Начальная сферическая симметрия поля и масштабная
инвариантность нарушаются когда размер вселенной вырастает до нескольких
единиц $r_\circ$, т.е. когла простейшие частицы, формирующие трипольные
структуры, переходят из нестабильного состояния с энергией $E_\circ$ и 
$R_Y=0$ 
в основное состояние
с равновесным радиусом $R_Y \approx 0.58\,r_\circ$.
Сферическая симметрия нарушается потому, что у каждого триполя 
появляется ось вращения, а масштабная инвариантность нарушается
из-за того, что компоненты триполя невозможно удалить 
из этой системы (см. \S\,4).

Дальнейший рост размеров вселенной приводит к формированию сложного
переплетения цепочек и кольцевых структур \cite{yershov05}. 
Эти структуры воздействуют стохастически друг на друга и 
на метрику пространства-времени, что с большой долей вероятности 
должно отразиться на последующем формировании крупномасштабной 
структуры вселенной точно также, как и в стандартной космологической
модели \cite{hagedorn65,omnes69}.

\section{Заключение.}

Как мы видим, предлагаемая модель дает возможность
решить проблему космологической сингулярности и
указать на источник происхождения наблюдаемого в природе 
набора элементарных частиц. 
Модель имеет встроенные единицы длины, времени и скорости, 
позволяющие обойтись без внешних или внутренних настраиваемых 
параметров.
В рамках этой модели динамику частиц и их компонентов можно
описывать в терминах стандартной релятивистской механики и
электродинамики Максвелла. Несмотря на то, что характерный 
масштаб расстояний модели вероятнее всего близок к пока 
экспериментально недостижимой планковской
длине, эту модель в принципе можно использовать для расчета параметров гораздо
б\'ольших структур (таких как, например, нуклоны). Поэтому 
интересной областью дальнейшего применения модели будет исследование 
ее феноменологических предсказаний в масштабах ядерных и 
атомных структур -- там, где эта модель может быть проверена
экспериментально уже сейчас.   

\vspace{0.2cm}
\noindent
Помимо проблем, упомянутых в предыдущем разделе,
модель хромоэлектрических преонов может пролить свет и на целый ряд 
других проблемных аспектов, в основном имеющих отношение к ядерной физике
и физике элементарных частиц, таких как  
неразличимость частиц, рождающихся в различных реакциях,
левополяризованность нейтрино,
механизм нарушения электрослабой симметрии,
происхождение принципа исключения Паули,
инвариантность масс и зарядов частиц 
и некоторых других.
Проблема происхождения наблюдаемого разнообразия частиц имеет отношение
к проблеме единственности вселенной \cite{susskind03}. 
Как отмечалось в \S\,5, в нашем случае конфигурация 
и число компонентов в каждой преонной структуре определены
единственным образом минимумом комбинированного потенциала 
этой структуры. Следовательно, наблюдаемое многообразие 
частиц, как и их внутренние структуры, определены
однозначно, и вселенная, смоделированная с использованием поля  
(\ref{eq:superpos}) будет генерировать всегда один и тот же 
набор частиц с одними и теми же свойствами. Важно заметить, 
что этот набор предопределен симметрией SU(3)$\times$U(1) поля,
так что выбор любой другой функциональной формы этого поля (отличающейся 
от, например, экспоненциальной формы, использованной в данной работе) приведет
лишь к изменению единиц масштаба, и не изменит собственно
состава набора частиц.    
Таким образом, в рамках модели хромоэлектрических преонов 
наблюдаемые нами свойства вселенной являются следствием 
трехмерности пространственных координат и одномерности временной 
координаты, что и определяет единственность вселенной.
В этом плане недостатком модели является невозможность
объяснения причин, которые приводят к трехмерности пространства.
Поэтому эту трехмерность приходится постулировать, основываясь
на экспериментальных данных.
Остальные аспекты, упомянутые выше, мы обсудим в других
статьях, отмечая здесь, что  
решение большинства проблемных вопросов физики частиц и космологии 
с использованием лишь одного-двух исходных принципов 
представляется весьма привлекательным.


\begin{thebibliography}{99}

{\small 
\bibitem{lemaitre33} 
G. Lema\^{\i}tre, 
\textit{Ann. Soc. Sci. Bruxelles A} {\bf 53}, 51--85 (1933);
%
R.\,C. Tolman, ``Relativity, Thermodynamics and Cosmology'' 
(Oxford University Press, Clarendon Press, Oxford, 1934). 
%
 
%
\bibitem{steinhardt02a} 
%
P.\,J. Steinhardt and N. Turok,  
\textit{Phys. Rev. D} {\bf 65}, 126003 (2002).
%
 
%
\bibitem{markov84}
M.\,A. Markov,  
\textit{Ann. Phys.}, {\bf 155}, 333--357 (1984);
%
S.\,A. Bludman,  
\textit{Nature}, {\bf 308}, 319--322 (1984);
%
R. Penrose,   \textit{Ann. N.Y. Acad. Sci.}, 
{\bf 571}, 249--264 (1989).
 

\bibitem{guth83}
A. Guth, M. Sher,   \textit{Nature},
{\bf 302}, 505--506 (1983);
%
J.\,.V. Narlikar, G. Burbidge, R.\,G. Vishwakarma,  
 \textit{J. Astrophys. Astr.}, {\bf 28}, 67--99 (2007).

%
\bibitem{wheeler62}
J.\,A. Wheeler, ``Geometrodynamics'' 
(Academic Press, New York, 1962). 
 
\bibitem{rovelli04}
%
C. Rovelli, ``Quantum Gravity'' (Cambridge Univ. Press, Cambridge, 2004).
%
 
\bibitem{markov82}
М.\,А. Марков,  
\textit{Письма ЖЭТФ}, {\bf 36}, 214--216 (1982); 
%
Э.\,Г. Аман, М.\,А. Марков, 
\textit{Теор. и мат. физика}, {\bf 58}, 163--168 (1984).
\bibitem{einstein35}
A. Einstein, N. Rosen,   \textit{Phys. Rev.}, {\bf 48}, 73--77 (1935).
 
%
\bibitem{finkelstein59}
D. Finkelstein and. C.\,W. Misner, 
  \textit{Ann. Phys.}
{\bf 6}, 230--243 (1959);
%
C.\,F.\,E. Holzhey and F. Wilczek,  
\textit{Nucl. Phys. B} {\bf 380}, 447--477 (1992);
%
V.\,V. Flambaum  and J.\,C. Berengut,  
\textit{Phys. Rev. D} {\bf 63}, 084010 (2001).
%
\bibitem{dsouza92}
I.\, A. D'Souza, C.\,S. Kalman, ``Preons'' (World Scientific, Singapore,
1992);
%
%
J.\,M. Gipson, Y. Tosa, R.\,E. Marshak,  
 \textit{Phys. Rev.} D, {\bf 32}, 284--292 (1985);
%
S. Fredriksson, arXiv:hep-ph/0309213.

\bibitem{pati74}
%
J.\,C. Pati, A. Salam,  
\textit{Phys. Rev.} D {\bf 10}, 275--289 (1974);
%
J.\,C. Pati, A. Salam,  \textit{Nucl. Phys.}, B {\bf 214}, 109--135
(1983);
%
%
J.-J. Dugne, S. Fredriksson and J. Hansson, \textit{Europhys. Lett.}, {\bf 60},
188--194 (2002).
 
\bibitem{artin26}
E. Artin,  
\textit{Abh. Math. Sem. Univ. Hamburg}, {\bf 4}, 174--177 (1926);
%
C.\,McA. Gordon,  \textit{Proc. Amer. Math. Soc.},
{\bf 58}, 361--362 (1976); 
%
C. Eling, T. Jacobson,   \textit{Class. Quant. Grav.}, {\bf 23},
5625--5642 (2006).
 
\bibitem{morris88}
M.\,S. Morris, K.\.S Thorne,  \textit{Am. J. Phys.}, {\bf 56}, 395--412 (1988);
%
C. Armendariz-Picon,  
\textit{Phys. Rev.} D, {\bf 65}, 104010 (2002).

\bibitem{shatskii08}
А.\,А. Шацкий, И.\,Д. Новиков, Н.\,С. Кардашев,
\textit{УФН}, {\bf 178}, 481--488 (2008). 
 
\bibitem{ritis83}
R. de\,Ritis, M. Lavorgna, G. Platania and C. Stornaiolo, 
  \textit{Phys. Rev. D} {\bf 28},
713--717 (1983);
%
M. Perry,   \textit{Phys. Lett.} B, 71, 234--236 (1977);
%
R.\,M. Kiehn,  
in: ``Topological Fluid Mechanics'', H.\,K.Moffatt and T.\,S. Tsinober, Eds.
(Cambridge Univ. Press, Cambridge, 1990) 449--458;
A. Dimakis and F. Muller-Hoissen,  
\textit{J. Math. Phys.} {\bf 44}, 1781--1821 (2003).
%

%
\bibitem{ivanenko38}
%
D.\,D. Iwanenko,  
\textit{Physikalische Zeitschrift der Sowjetunion}, {\bf 13}, 141--150 (1938);
%
W. Heisenberg,  
 \textit{Nachr. Akad. Wiss. G\"ottingen}, IIa, 111--127 (1953);
%
V.\,D. Dzhunushaliev,   \textit{Intern. J. Mod. Phys. D}
{\bf 7}, 909--915 (1998);
%
V.\,G. Krechet and V.\,N. Ponomarev,  
\textit{Theor. Math. Phys.}, {\bf 25}, 1036--1038 (1975).
%
%
\bibitem{jackiw00}
R. Jackiw, V.\,P. Nair and  S.-Y. Pi, 
 \textit{Phys. Rev. D} {\bf 62}, 085018 (2000);
 
J. Dai and V.\,P. Nair,  
\textit{Phys. Rev. D} {\bf 74}, 085014 (2006).
 
\bibitem{suisso02} 
 E.\,F. Suisso, J.\,P.\,B.\,C. de Melo and T. Frederico,  
 \textit{Phys. Rev. D} {\bf 65},
094009 (2002).
 
\bibitem{wiggins89}
S.\,Wiggins, ``Global Bifurcations and Dynamical Chaos'' (Springer-Verlag, Berlin, 1988);
%
L.\, Carlson and W.\,C.\,Schieve,  
 \textit{Phys. Rev.} A {\bf 40}, 1127--1129 (1989).
%

 
\bibitem{pati81}
J.\,C. Pati,  
\textit{Phys. Lett.} B, {\bf 98}, 40--44 (1981).
%
\bibitem{kleman83} 
M.\,Kleman, ``Lines and Walls'' (Wiley and Sons, Chichester, 1983).
%
\bibitem{yershov07}
V.\,N. Yershov, \textit{Physica D} {\bf 226}, 136--143 (2007).

\bibitem{yershov05} 
V.\,N. Yershov,  
\textit{Few-Body Syst.} {\bf 37}, 79--106 (2005).

\bibitem{gliner75}
E.\,B. Gliner, I.\,G. Dymnikova,  
\textit{Sov. Astron. Lett.}, {\bf 1}, 93--94 (1975).

\bibitem{markov83}
M.\,A. Markov,  \textit{Phys. Lett.} A, {\bf 94}, 427--429 (1983).

\bibitem{penrose82}
R. Penrose,  in \textit{Progress in Cosmology}, Ed. A.\,W. Wolfendale
(D. Reidel Publ. Comp., Dordrecht 1982) 87--88.

\bibitem{penrose89}
R. Penrose, ``The Emperor's New Mind'' (Oxford Univ. Press, Oxford 1989);
R. Penrose, ``The Road to Reality'' (Vintage Books, 2006).

\bibitem{tolman31}
R.\,C. Tolman,  
\textit{Phys. Rev.}, {\bf 37}, 1639--1660 (1931).

\bibitem{brown07}
L. Brown, P.\,H. Frampton,  
\textit{Phys. Rev. Lett.}, {\bf 98}, 071301 (2007).

\bibitem{lahav04}
O. Lahav, A.\,R. Linde, arXiv:astro-ph/0406681;
%
A. Blanchard, arXiv:astro-ph/0301137;
%
K. Lake,  \textit{Phys. Rev. Lett.},
{\bf 94}, 201102 (2005).
 

\bibitem{penrose79}
R. Penrose,  
in General Relativity: An Einstein Centenary Survey, ed. S. Hawking
and W. Israel (Cambridge Univ. Press, Cambridge, 1979) 581--638;
%
%
M.\,M. \'Cirkovi\'c,   
\textit{Found. Phys.} {\bf 32}, 1141--1157 (2002);
%
Ya.\,B. Zeldovich,  
 \textit{Mon. Not. R. Astr. Soc.}, {\bf 160}, 1--3 (1972).
%

\bibitem{penrose90}
R. Penrose,  
 \textit{Ann. N.Y. Acad. Sci.}, 
{\bf 571}, 249--264 (1989).

\bibitem{zeldovich75}
Я.\,Б. Зельдович, И.\,Д. Новиков, ``Строение и эволюция вселенной'',
(Москва, \glqqНаука\grqq 1975), Гл.\,23 \S\,9.

\bibitem{lifschitz63}
E.\,M. Lifschitz, I.\,M. Khalatnikov,  
\textit{Adv. Phys.}, {\bf 12}, 207 (1963).

\bibitem{penrose69}
R. Penrose,  
\textit{Phys. Rev. Lett.}, {\bf 14}, 57--59 (1965);
%
S.\,W. Hawking,   \textit{Proc. R. Soc. London} A 
{\bf 300}, 187--201 (1967);
%
R. Penrose,  
\textit{Riv. Nuov. Cim.}, {\bf 1}, 252--276 (1969);
%
S.\,W. Hawking and R. Penrose,  
 \textit{Proc. R. Soc. Lond. A} {\bf 314}, 529--548 (1970);
%
S.\,W. Hawking, G.\,F.\,R. Ellis, ``The Large Scale Structure of Space-Time''
(Cambridge Univ. Press, Cambridge, 1973).

\bibitem{tangherlini93}
F.\, R. Tangherlini,  
\textit{Il Nuovo Cim.} B, {\bf 108}, 1253--1273 (1993).

\bibitem{durrer96}
R. Durrer, J. Laukeumann,  \textit{Class. Quant. Grav.}, {\bf 13}, 1069--1088
(1996).

%
\bibitem{thurston84}
W.\,P. Thurston and  J. Weeks,  
\textit{Sci. Am.} {\bf 251}, 108--113 (1984);
%
W.\,P. Thurston, ``Three-Dimensional Geometry and Topology'' (Princeton Univ. Press.,
Princeton, 1997);
%
M. Anderson,  
\textit{Ann. Scient. Ecole Norm. Sup.} {\bf 18}, 89-105 (1985).
%


\bibitem{jakobson06}
D. Jakobson, N. Nadirashvili and I. Polterovich,  
\textit{Canad. J. Math.}
{\bf 58}, 381--400 (2006). 
 
\bibitem{spergel03}
D.\,N. Spergel et al,  
\textit{Astrophys. J. Suppl.} {\bf 148}, 175--194 (2003). 

%
\bibitem{albrecht99}
%
A. Albrecht, J. Magueijo,  
\textit{Phys. Rev.} D{\bf 59}, 043516 (1999);
%
%
M. Szydlowski and A. Krawiec A.,  arXiv:\,gr-qc/0212068;
%
G. Ellis, R. Maartens,  arXiv:\,gr-qc/0211082;
%
J.\,C. Niemeyer,   \textit{Phys. Rev. D} {\bf 65}, 083505 (2002);
%
J. Magueijo, 
\textit{Phys. Rev. D} {\bf 63}, 043502 (2001).
%

\bibitem{chakraborty02}
S. Chakraborty,  
\textit{Nuovo Cimento B} {\bf 117}, 189--195 (2002).

\bibitem{goedel49}
K. G\"odel,   \textit{Rev. Mod. Phys.}
{\bf 21}, 447--450 (1949);
%
K. Lanczos,  
\textit{Zeitschr. f. Phys.}, {\bf 21}, 73 (1924);
%
G. Gamow,  \textit{Nature}, {\bf 158}, 549 (1946); 
%
A. Raychaudhuri,  \textit{Phys. Rev.},
{\bf 98}, 1123--1126 (1955);
%
C.\,B. Collins, S.\,W. Hawking,  
\textit{Mon. Not. R. Astron. Soc.}, {\bf 162}, 307--320 (1973).
%
Yu.\,N. Obukhov,   \textit{Gen. Rel. Grav.},
{\bf 24}, 121--128 (1992).


\bibitem{fan50}
C.\,Y. Fan,  \textit{Phys. Rev.}, {\bf 77},
140 (1950).

\bibitem{magueijo03} 
J. Magueijo, 
  \textit{Rept. Prog. Phys.} 
{\bf 66}, 2025--2068 (2003).
%

\bibitem{hagedorn65} 
R. Hagedorn,  
\textit{Nuovo. Cim. Suppl.}, {\bf 3}, 147--186 (1965);
%
R. Hagedorn,  
\textit{Z. Phys. C: Par. Field.}, {\bf 17}, 265--281 (1983).

%
\bibitem{omnes69}
R. Omn\`es,  
\textit{Phys. Rev. Lett.}, {\bf 23}, 38--40 (1969);
%
%
Ya.\,B. Zeldovich,  
 \textit{Mon. Not. R. Astr. Soc.} {\bf 192}, 663--667 (1980);
%
A.\,A. Starobinsky,  
\textit{Phys. Lett.} B {\bf 91}, 99--102 (1980);
%
L. Rezzolla, J.\,C. Miller, and O. Pantano,  
\textit{Phys. Rev.} D, {\bf 52}, 3202--3213, (1995).
%
%
P.\,C. Fragile,  P. Anninos,  
 \textit{Phys. Rev.} D, {\bf 67}, 103010, (2003).
%
 

\bibitem{susskind03}
L. Susskind,
arXiv:hep-th/0302219; G.\,F.\,R. Ellis,  
arXiv:astro-ph/0602280.
}
\end{thebibliography}
\end{document}